\documentclass[5p,twocolumn]{elsarticle}
\usepackage{graphicx,latexsym}
\usepackage{dcolumn}
\usepackage{amssymb,amsmath,bm}
\usepackage{subfigure}
\usepackage{braket}
\usepackage{siunitx}
\usepackage{array}
\usepackage{float}
\usepackage[nopar]{lipsum}
\usepackage{caption}
\usepackage{multirow}
\usepackage{bigstrut}

\biboptions{sort&compress}

\usepackage[super]{nth}

\newcommand{\angstrom}{\text{\normalfont\AA}}
\usepackage{hyperref}
\hypersetup{
    pdfnewwindow=true,       
    colorlinks=true,         
    linkcolor=blue,          
    citecolor=blue,          
    filecolor=magenta,       
    urlcolor=black           
}

\usepackage[normalem]{ulem}

\def\sec#1{Sec.\ \ref{#1}}

\def\fig#1{Fig.\ \ref{#1}}
\def\tab#1{Tab.\ \ref{#1}}

\journal{}

\begin{document}

\begin{frontmatter}


\title{Optical conductivity enhancement and thermal reduction of BN-codoped MgO nanosheet: Significant effects of B-N atomic interaction}

\author[a1,a2]{Nzar Rauf Abdullah}
\ead{nzar.r.abdullah@gmail.com}
\address[a1]{Division of Computational Nanoscience, Physics Department, College of Science,
             \\ University of Sulaimani, Sulaimani 46001, Kurdistan Region, Iraq}
\address[a2]{Computer Engineering Department, College of Engineering,
	\\ Komar University of Science and Technology, Sulaimani 46001, Kurdistan Region, Iraq}

\author[a3]{Botan Jawdat Abdullah}
\address[a3]{Physics Department, College of Science- Salahaddin University-Erbil, Erbil 44001, Kurdistan Region, Iraq}

\author[a4]{Yousif Hussein Azeez}
\address[a4]{Physics Department, College of Science, University of Halabja, Kurdistan Region, Iraq.}

\author[a5]{Chi-Shung Tang}
\address[a5]{Department of Mechanical Engineering,
	National United University, 1, Lienda, Miaoli 36003, Taiwan}

\author[a6]{Vidar Gudmundsson}
\address[a6]{Science Institute, University of Iceland, Dunhaga 3, IS-107 Reykjavik, Iceland}


\begin{abstract}

We investigate the electronic, the thermal, and the optical properties of BN-codoped MgO monolayers taking into account the interaction effects between the B and the N dopant atoms. The relatively wide indirect band gap of a pure MgO nanosheet can be changed to a narrow direct band gap by tuning the B-N attractive interaction.
The band gap reduction does not only enhance the optical properties, including the absorption spectra and the optical conductivity, but also the most intense peak is shifted from the Deep-UV to the visible light region. The red shifting of the absorption spectra and the optical conductivity are caused by the attractive interaction. In addition, both isotropic and anisotropic characteristics are seen in the optical properties depending on the strength of the B-N attractive interaction.
The heat capacity is reduced for the BN-doped MgO monolayer, which can be referred to changes in the  bond dissociation energy. The bond dissociation energy decreases
as the difference in the electronegativities of the bonded
atoms decreases. The lower difference in the electronegativities leads to a weaker endothermic process resulting in reduction of the heat capacity. An ab initio molecular dynamics, AIMD, calculation is utilized to check the ther­modynamic stability of the pure and the BN-codoped MgO monolayers.
We thus confirm that the BN-codopant atoms can be used to gain control of the properties of MgO monolayers for thermo- and opto-electronic devices.

\end{abstract}


\end{frontmatter}

\section{Introduction}

Low-dimensional nanomaterials have different unique characteristics and have made significant contributions to the progress of nanotechnology. Very thin nanomaterials with unique physical characteristics are known as two-dimensional (2D) materials. Due to their distinctive physical characteristics and remarkable capabilities, the 2D materials have attracted a great deal of attention since the discovery of graphene \cite{novoselov2004electric}. The properties of many 2D materials have been modified in unique ways, leading to even more extraordinary results in terms of regulated factors and functioning.
This has prompted scientists to expand their studies on 2D materials. Due to their unique electrical, optical, thermal, magnetic, mechanical, and catalytic capabilities, 2D materials are essential in nanotechnology and demand more extensive research \cite{shayeganfar2017monolayer, akinwande2017review, novoselov2005two, xiao2019atomic, abdullah2021properties, abdullah2022electronic}.

A variety of new two-dimensional materials are being researched for their unique features in the context of the discovery of graphene. It is possible to successfully synthesize other 2D materials with qualities that are approximately at level with or better than the properties of graphene \cite{mak2016photonics, sangwan2013low, vogt2012silicene, ABDULLAH2022140235, mannix2018borophene, bianco2013stability, zhu2015epitaxial}. Experimentally, synthetic materials are produced in different of ways and MgO nanosheets are synthesized using methods such as layer-by-layer epitaxial growth mode of MgO films by pulsed laser deposition \cite{matsuzaki2010layer, C6NR08586E}. A synthesizing sol-gel process has been used to create MgO nanosheets \cite{li2011experimental, kamarulzaman2016band, thomele2021cubes}, MgO is produced through the thermal and pressure synthesis of methanol and ethanol from a brown coal fly ash waste \cite{zhu2012preparation, qian2021synthesis, liu2021ultrathin}, and MgO nanosheets are formed using a wet chemical process via chemical vapor deposition \cite{zhao2017generalized, wang2014mgo}.

A detailed study of several monolayers of the II-VI semiconductor family with characteristics like 2D graphene has been carried out using first principles methods. Dynamically stable hexagonal monolayers have been found for some of them, such as MgO \cite{zheng2015monolayer, luo2019graphene}. The MgO monolayers have a number of intriguing applications due to unusual surface activity, including MgO(111) for methanol-based fuel cells \cite{hu2007mgo}. They have the capacity to cause uncommon reactions \cite{liu2021atomically}. They show catalytic activity for the toxic adsorption of a number of environmental pollutants \cite{zhu2006efficient}. They are used for carbon dioxide and methane reforming catalytic properties \cite{lin2015effect}, for ion batteries as anode materials for producing ultra-long-term reversible rechargeable batteries with high energy density and capacity retention at a high current density \cite{zhou2020redistributing}, and have been used as an ideal substrate \cite{smerieri2015spontaneous, zhang2014carbon}.

Although there are many theoretical studies on MgO nanosheets for some limited parameters identified, it is still fascinating to explore them further. For examples, the structure, the stability, and the electrical properties of MgO nanosheets have been investigated using first principles calculations, and it has been shown that a MgO nanosheet has semiconductive properties with band gaps varying from 4.23 to 4.38 eV with different applied strain \cite{zhang2012structural}. The calculated band gap for MgO monolayers with a direct band gap is 3.1 eV for the GGA method and 4.2 eV for GGA-mBJ DFT approach, according to studies of its electrical and optical properties based on first principles. Additionally, an RPA approach used to determine some optical parameters shows the semiconducting properties of a MgO monolayer \cite{nourozi2019electronic}. The phonon and the thermodynamic properties of MgO (111) and MgO (100) nanosheets have been studied by first-principle calculations. Due to the fact that MgO (100) exhibits stronger dynamical stability than MgO (111) for the nanosheet gradient of the acoustic branches, the ensuing phonon dispersion curves suggest that MgO has a higher thermal conductivity  \cite{yeganeh2019vibrational}, and quantum confinement's effects on the thermoelectric and electrical properties of bulk MgO monolayers, MgO(111), and MgO(100) monolayers have been studied using DFT and Boltzmann models \cite{asl2022two}.

Additionally, DFT has been used to examine the electrical and optical characteristics of an MgO nanosheet under compressive and tensile strain. According to the results, the band gap energy is 3.45 eV and decreases under tensile strain, but a rising trend is shown when compressive strain is used \cite{yeganeh2019effects}. In general, doping is an issue for understanding and modifying the monolayer properties. Theoretical investigations have been carried out on doped MgO monolayer by one element, which indicate the possibility of future distinguishing features. Theoretical research on MgO monolayers is now dedicated to certain doping elements using DFT such as: the electrical and magnetic properties of a transition metal doped MgO monolayer have been investigated, and the results reveal that the band gaps and the magnetic characteristics of MgO sheets can be altered \cite{wu2014electronic}. The electrical and the magnetic properties of MgO monolayer with B, C, N, and F dopant atoms have been studied, and the results show that due to the appearance of impurity states within the band gap, the band gaps of doped MgO monolayers can be tuned, whereas F-doped MgO monolayers achieve the transition from a semiconductor to a metal. Furthermore, doped MgO monolayers display magnetic properties as a result of the polarizations of the dopants and nearby Mg or O atoms, however no magnetism is seen in the case of F doped MgO monolayers \cite{wu2016first}.

The electrical, optical and magnetic properties of a MgO monolayer with B, C, N, and F dopant atoms have been studied. According to the electronic structure calculations, it is found that a MgO monolayer is a semiconductor with an indirect band gap of approximately 3.8 eV at the gamma point. Additionally, various magnetic properties are obtained, such that the B and C doped MgO monolayers have magnetic half-metal properties, while the N doped MgO is an antiferromagnetic semiconductor, and F doped MgO has a nonmagnetic metal property. Both the MgO and F-doped MgO monolayers are transparent according to optical estimates \cite{moghadam2018electronic}.
The impact of point defects and doping by Si and Ge on the electrical and magnetic characteristics of the MgO monolayer has been investigated finding the band gap of the MgO monolayer to be 3.37 eV. The degree to which the electronic band gap fluctuates under certain circumstances can cause the material to transform from being a non-magnetic material to a ferromagnetic semiconductor, and similar to this, non-magnetic characteristics are obtained when one Mg atom is replaced with one Si or Ge atom  \cite{van2022defective}.

Monolayers of MgO have only undergone a limited number of doping investigations, particularly for codopants for the identification of differentiating features. Therefore, this study uses first-principles computations to evaluate the electrical, the thermal, and the optical characteristics of a monolayer MgO doped with B or N searching for modified physical parameters. The BN codoping is studied for the first time. The findings show that the B and the N codopants significantly affect the MgO monolayer's band structures, reducing the energy gap while preserving the structure's semiconductor properties. The outcomes additionally demonstrate an improvement in the electrical, the thermal, and the optical properties. This means that the BN-codoped MgO monolayer can be significant for applications in thermoelectric and optoelectronic technologies.

The paper is organized as follows: Details of the computational techniques are included in the \sec{methodology}. The calculated electrical, thermal, and optical properties for a pure MgO monolayer and with BN codopants are presented in \sec{results}. The conclusion is included in \sec{conclusion}.

\begin{table*}[h]
	\centering
	\begin{center}
		\caption{\label{table_one} The average bond lengths of Mg-O, Mg-B, Mg-N, and B-O, the average lattice constant, $a$, the formation energy (E$_{f}$), and the band gap, E$_{\rm g}$.
		The unit of bond lengths and lattice constant is $\angstrom$, and the formation energy and the and gap is $eV$.}
		\begin{tabular}{l|l|l|l|l|l|l|l}\hline
			Structure&  Mg-O      &  Mg-B     &  Mg-N    & B-O  & $a$    &  E$_f$   &  E$_{\rm g}$   \\ \hline
			MgO	     &  1.90      &  -        &  -       &  -   &  3.29  &   -11.7  &   3.4     \\
			B-MgO    &  1.944     &  2.215    &  -       &  -   &  3.53  &   -10.6  &    -      \\
			N-MgO    &  1.916     &  -        &  1.99    &  -   &  3.36  &   -10.9  &    -      \\
			BN-MgO-I &  1.895     &  -        &  1.96    & 1.49 &  3.013 &   -8.59  &   1.7     \\
			BN-MgO-II&  2.03      &  -        &  1.81    & 1.43 &  3.038 &   -9.85  &   0.54    \\ \hline
	\end{tabular}	\end{center}
\end{table*}

\section{Methodology}\label{methodology}

We assume a 2D hexagonal MgO, h-MgO, monolayer structure consisting of a $2\times 2$ supercell with equal number of Mg and O atoms. In addition, B, N, and B-N atoms are used as dopant to tune the physical properties of MgO monolayer.
The inter-layer interaction of the MgO monolayers is canceled out by considering $20 \, \angstrom$ of vacuum between the layers in the $z$-direction.
All the first principle calculations for the structure optimization, the electronic, and the optical properties are carried out within the framework of density functional theory using the Quantum espresso, QE, package \cite{Giannozzi_2009, giannozzi2017advanced}.
The pure and doped MgO monolayers are fully relaxed with cutoffs for the kinetic energy and the charge densities at $1088.5$~eV, and $1.088 \times 10^{4}$~eV, respectively, and a $18\times18\times1$ $k$-mesh is used to sample of the primitive cell.
All crystal structures are relaxed until all the forces on the atoms are less than $10^{-5}$ eV/$\angstrom$.
The generalized gradient approximation (GGA) with the Perdew-Burke-Ernzerhof (PBE) functional is used for the exchange-correlation potential. In the density of states calculations, we use a $100 \times 100 \times 1$ $k$-mesh grid \cite{ABDULLAH2022106943, ABDULLAH2023116147}.
The optical properties are obtained using QE with the optical broadening of $0.1$~eV.
The thermal calculations are done using the dmol$^3$ package based on a DFT formalism  \cite{doi:10.1063/1.458452}.

An ab initio molecular dynamics, AIMD, calculations are utilized to check the ther­modynamic stability. The calculations, done in the NVT ensemble, are performed for $10$~ps with a time step of $1.0$~fs using the heat bath approach described by Nosé-Hoover \cite{doi:10.1063/1.463940}.

\section{Results}\label{results}

To study the properties of a BN-doped 2D h-MgO monolayer, the structural, electronic, thermal, and optical properties of a pure h-MgO monolayer are computed for the sake of comparison to the BN-doped MgO.

\subsection{Structural properties}

\begin{figure}[htb]
	\centering
	\includegraphics[width=0.35\textwidth]{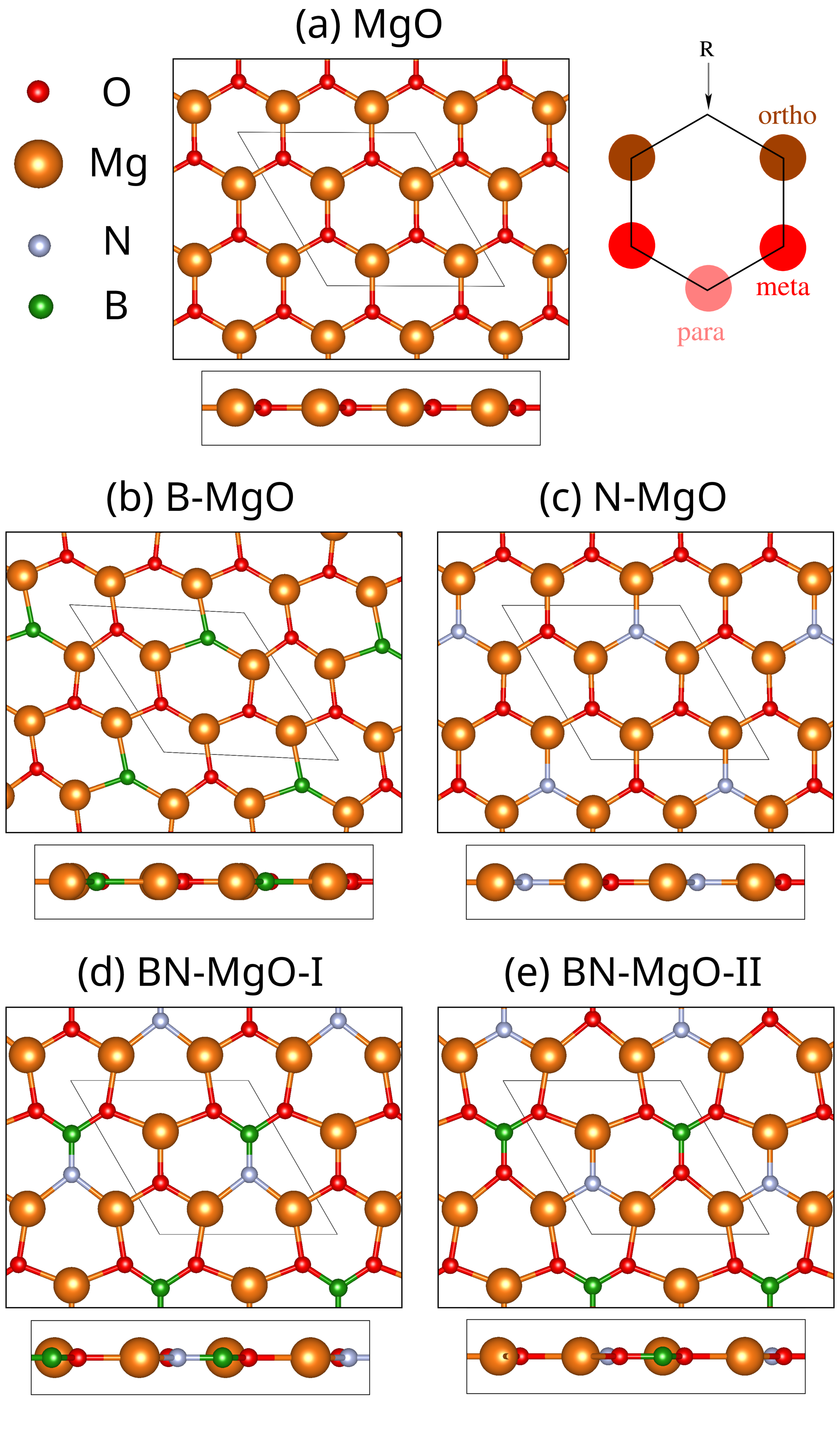}
	\caption{Crystal structure of pure MgO, and B-, N-doped, and BN codoped MgO monolayers. The side view of each structure is shown in its bottom panel.}
	\label{fig01}
\end{figure}

We consider a pure MgO monolayer, a B- or an N-doped MgO, and a BN-codoped MgO monolayer that are displayed in \fig{fig01}. The B (b) or the N (c) are substitutionaly doped at the position of the O atoms,
while the B or N substitutionaly doped at the Mg position is not stable (not shown). If the B- or N-atoms are doped in the Mg position, a large deformation in the structure is seen, which is due to a large difference in the atomic radius of B- or N-atoms comparing to the Mg atoms.

There are two stable atomic configurations of BN-codoped MgO shown in \fig{fig01}(d,e), which
are identified as BN-MgO-I (d), and BN-MgO-II (e). In the BN-MgO-I configuration, a B(N) atom is doped at the ortho(meta) position in the hexagonal honeycomb structure forming the B-N bond with length of $1.416$~$\angstrom$. The B-N bond is not formed in the BN-MgO-II where the B atom is doped at the ortho position and the N atom is at meta position of the other side of hexagon leading to a B-N distance with length of $3.62$~$\angstrom$. It is clear that the B and the N atoms in BN-MgO-I and BN-MgO-II are doped at the same atomic position with different distance between the B and the N, which will play an important role in the interaction between the B and the N atoms in the structures.

The atoms in a structure interact with one another and this type of interaction is called
the inter-atomic interaction forming energy.
The interaction energy of each considered structure can be calculated from the total energy \cite{RASHID2019102625}. The interaction energy between
the B and the N atoms, ( $\Delta \rm E^{BN}$ ), can be introduced as the difference between
the energy of the dimer ($\rm E^{BN}$) and the sum of the monomer energies ($\rm E^{B} + E^{N}$) \cite{FALKOVSKY20085189}. We thus find the interaction energy between the dopant atoms, the B and the N atoms, in both BN-MgO-I and BN-MgO-II with values $-1.82$ and $-2.14$~eV, respectively. The negative sign of the interaction energy indicates that the interaction is attractive, and the attractive interaction in BN-MgO-I is larger than that of BN-MgO-II. This is due to the smaller distance between the B and the N atoms in BN-MgO-I compared to BN-MgO-II.

The interaction energy influences the structural properties of the monolayers including the average bond length and the average lattice constant as is presented in \tab{table_one}. The lattice constant, $a$, of pure MgO is $3.29$~$\angstrom$, but is increased to $3.53$, and $3.36$~$\angstrom$ in the presence of B (B-MgO), and N (N-MgO), respectively. This could be expected as the atomic radius of the B and the N atoms is larger or comparable to the O atoms. The increased value of $a$ indicates that these two structures are expanded.
In contrast, the lattice constant in BN-MgO-I and BN-MgO-II are decreased due to the attractive interaction between the B and the N atoms. A compressed super-cell is thus seen here.

Another item to study is the structural stability of the monolayers, which can be deduced from the formation energy, E$_f$, shown in \tab{table_one}. As is seen the pure MgO has the formation energy of -$11.7$~eV, which is very well in agreement with the literature \cite{doi:10.1021/jp3077062}. The formation energy is increased with the dopant atoms in which the largest formation energy is -$8.59$ for the BN-MgO-I. This is expected as the interaction energy is high in this monolayer, which leads to a high formation energy, but on the other hand it decreases the structural stability of the structure. In general, we can confirm that the pure and doped MgO structures are relatively stable.

\subsection{Electronic properties}

The electronic characteristics, such as band structure, the DOS, and the PDOS are presented here.
As was mentioned before, the GGA-PBE method is used to calculate the band structure which
leads the band gap of the monolayers to be underestimated \cite{ABDULLAH2022106835}.
\begin{figure}[htb]
	\centering
	\includegraphics[width=0.35\textwidth]{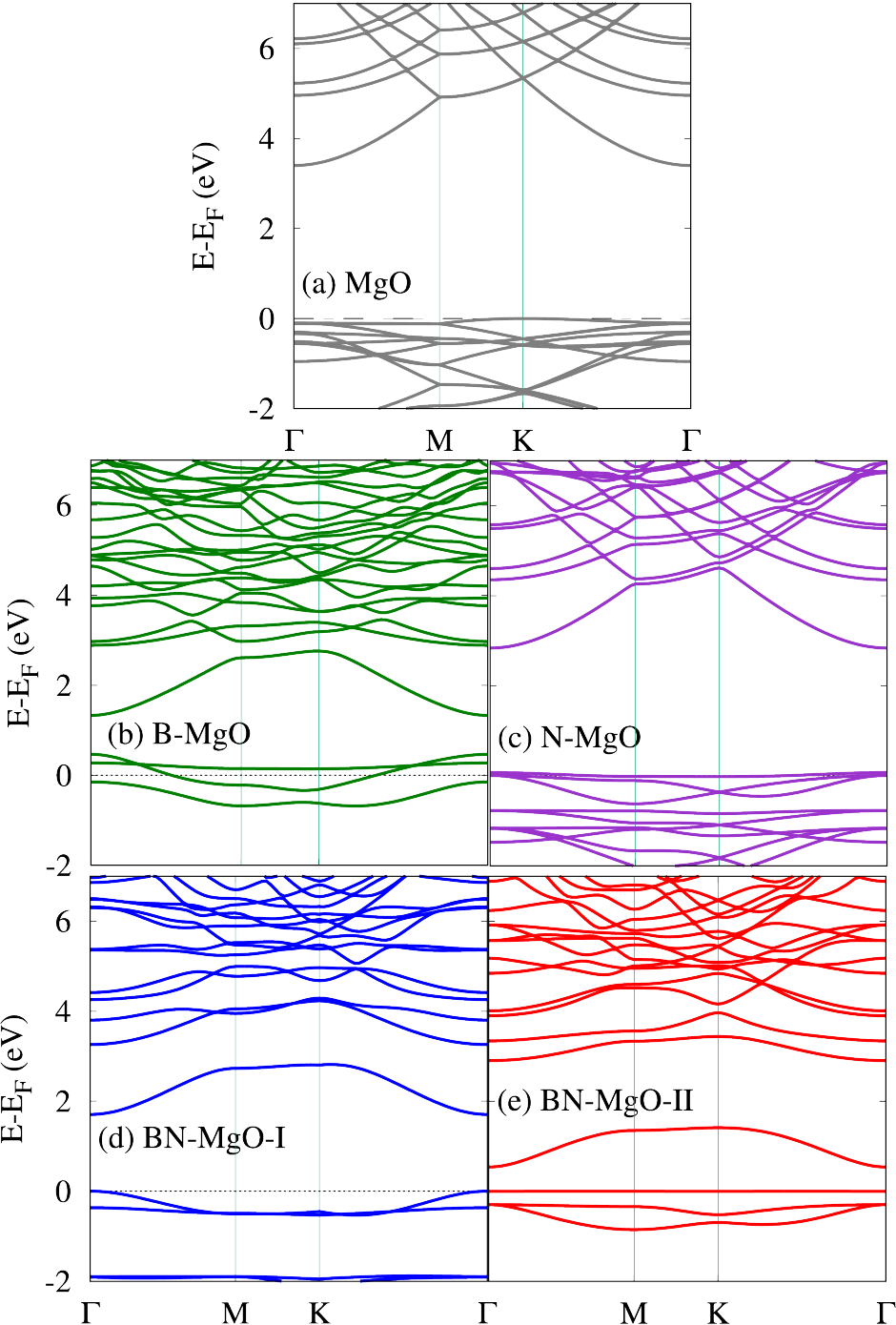}
	\caption{Band structure of pure MgO (a), B-MgO (b), N-MgO (c), BN-MgO-I (d), and BN-MgO-II (e).
		The energies are with respect to the Fermi level, and the Fermi energy is set to zero.}
	\label{fig02}
\end{figure}
Our DFT calculations demonstrate the indirect bandgap of pure MgO with a value of $3.4$~eV displaced from the K- to the $\Gamma$-point, which is very well in agreement with previous DFT calculations \cite{doi:10.1021/jp3077062}.
\begin{figure*}[htb]
	\centering
	\includegraphics[width=0.8\textwidth]{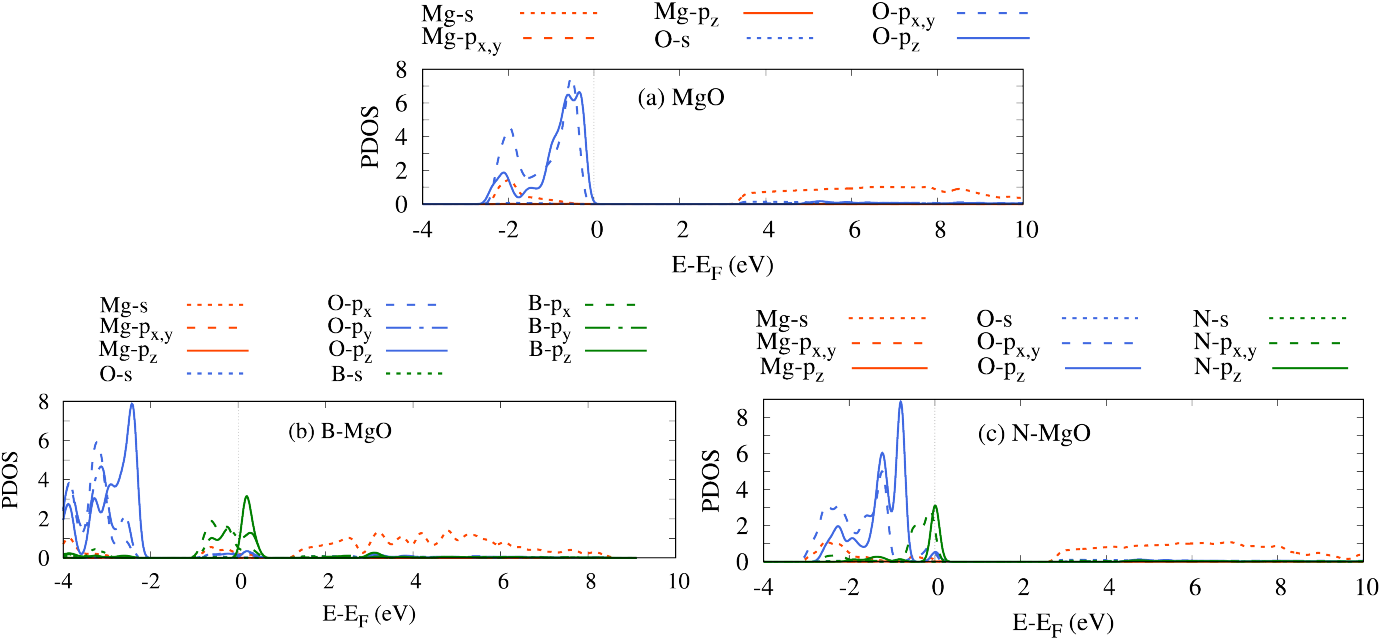}
	\caption{Partial density of state, PDOS, for pure MgO (a), B-MgO (b), and N-MgO (c).}
	\label{fig03}
\end{figure*}
\begin{figure*}[h!]
	\centering
	\includegraphics[width=0.8\textwidth]{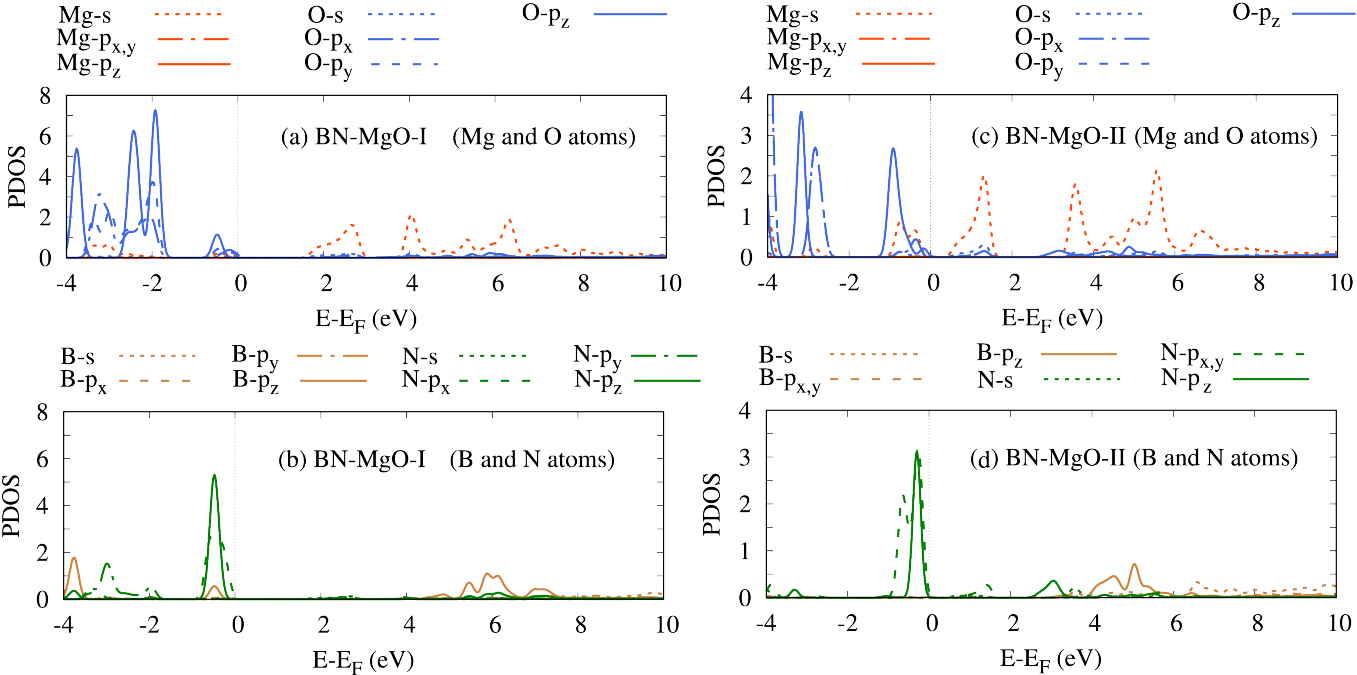}
	\caption{PDOS of BN-MgO-I (a,b) and BN-MgO-II (c,d) monolayers.}
	\label{fig04}
\end{figure*}
The wide band gap of a pure MgO monolayer indicates semiconducting characteristics as has been reported earlier \cite{C6NR08586E, doi:10.1073/pnas.1906510116, PhysRevB.92.115307}.
The B and the N dopant atoms in MgO have opposite influence on the band structure and the band gap of a pure MgO monolayer. A B atom shifts the conduction band down crossing the Fermi energy to the conduction band region, while an N doped MgO leads to the crossing of the Fermi energy into the valence band region as is seen in \fig{fig02}(b,c). In both cases, the monolayers have metallic properties. The energy crossing properties are seen in B- or N-doped graphene \cite{ABDULLAH2020126350} and BeO monolayers. It has been reported that B- or N-doped MgO could have both metallic and semiconductor properties based on the spin-component \cite{MOGHADAM201833}. The B-doped MgO has metallic and indirect semiconductor character at $\Gamma$-point for spin-up and down, respectively, while the N-doped MgO has direct band gaps about $3.2$~eV and $0.85$~eV exhibiting semiconductor characteristics in the spin-up and the spin-down channels. In our calculations, the spin effect is not taken into account, and a metallic character for both B- or N-doped MgO is thus found.

In contrast, a balance of energy crossing is found when BN-codoped MgO is considered in BN-MgO-I (d), and BN-MgO-II (e) of \fig{fig02}. In these two cases the energy crossing is not seen, instead, a band gap reduction is found and the Fermi energy remains between the valence and the conduction bands indicating semiconductor properties. The band gap is changed from indirect to direct in the cases of BN-codoped MgO, and the band gap of BN-MgO-I and BN-MgO-II are found to be $1.7$ and $0.54$~eV, respectively. The attractive interaction between the B and the N atoms plays an essential role in controlling the band gap. We mentioned that the B atoms shift the conduction bands down and the N atoms shift the valence bands up. The strong attractive interaction in BN-MgO-I will reduce the effects of of the energy shifts resulting in the opening up of the band gap. If the attractive force is reduced as is seen in BN-MgO-II, the shifting effect is increased and the band gap is thus further reduced \cite{ABDULLAH2022115554}.

In order to analyze the effects of doping, better understanding the nature of band structure, and  the orbital contribution to the conduction and valence bands,
the partial density of states, PDOS, for MgO (a), B-MgO (b), and N-MgO (c) are shown in \fig{fig03},
and the PDOS for BN-MgO-I (a,b) and BN-MgO-II (c,d) are displayed in \fig{fig04}.
It is found that the valence band maxima (VBM) and the conduction band minima (CBM) for a pure MgO monolayer arise from the $2p$ state ($p_{x,y}$ and $p_z$) of the O atoms and the $3s$ state of Mg atoms, respectively, which is in line with previous DFT observations \cite{D0RA05030J}.

In a B-MgO monolayer, the $p_x$-, $p_y$-, and $p_z$-states of B atoms make the main contribution to the crossing of the valence bands to the Fermi energy in addition to a small contribution of the $p_x$ and $p_z$ states of the O atom. So, we can confirm that the energy crossing is mainly due to the impurity atoms, the B atoms. In the N-MgO monolayer, the contributions of the above mentioned states of the N and the O atoms cross the Fermi energy leading to a metallic property. The difference here is that the $p_x$- and $p_y$-orbitals of the B and the O atoms are anti-symmetric in the B-MgO, while the $p_x$- and $p_y$-orbitals of the N and the O atoms are symmetric in the N-MgO monolayer.

In the BN-codoped MgO monolayers, the PDOS of Mg and O (a,c), and B and N (b,d) are shown in \fig{fig04} for BN-MgO-I and  BN-MgO-II monolayers. We find that the valence band maxima are mostly formed due to
the $p_z$ and $p_x$ of the impurity N atoms with a small contribution of $p_z$ of the B and the O atoms.
In addition, we see that the dopant atoms do not have an important role in tuning the conduction band maxima as most of the contribution is due to the $s$-orbitals of the Mg atoms. The $p_x$ and $p_y$ orbitals of the B and the N atoms in the BN-MgO-I are anti-symmetric, while in BN-MgO-II they are symmetric.

\subsection{Thermal properties}

\begin{figure}[htb]
	\centering
	\includegraphics[width=0.45\textwidth]{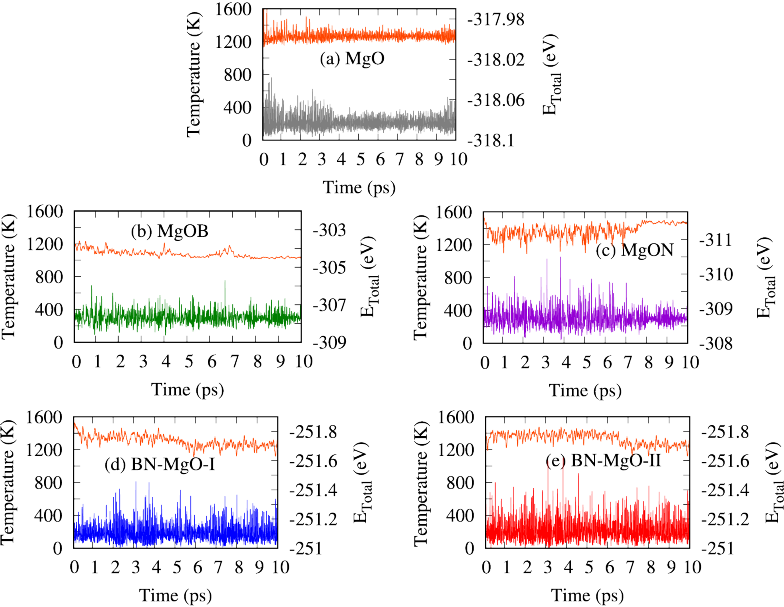}
	\caption{Temperature versus the AIMD simulation time steps at 300 K for optimized pure MgO (a), B-MgO (b), N-MgO (c), BN-MgO-I (d), and BN-MgO-II (e).}
	\label{fig05}
\end{figure}

The thermal stability of a pure and doped MgO is checked for approximately $10$ ps with a time step of $1.0$~fs as is shown in \fig{fig05}. The temperature curve of the pure and the doped MgO monolayers neither displays large fluctuations in the temperature nor serious structure disruptions or bond breaking at $300$~K. This indicates that the pure and the doped MgO monolayers are thermodynamically stable structures. In addition, the total energy (right side of $y$-axis) versus temperature is also presented (orange color), and the total energy also indicates no big fluctuations with respect to the temperature.

The thermal properties for pure and BN-codoped MgO monolayers are presented here.
The heat capacity is the ratio of the heat absorbed by a material to the temperature change.
It displays not only the thermal energy
stored within a monolayer, but also how quickly the monolayer cools or heats.

\begin{figure}[htb]
	\centering
	\includegraphics[width=0.4\textwidth]{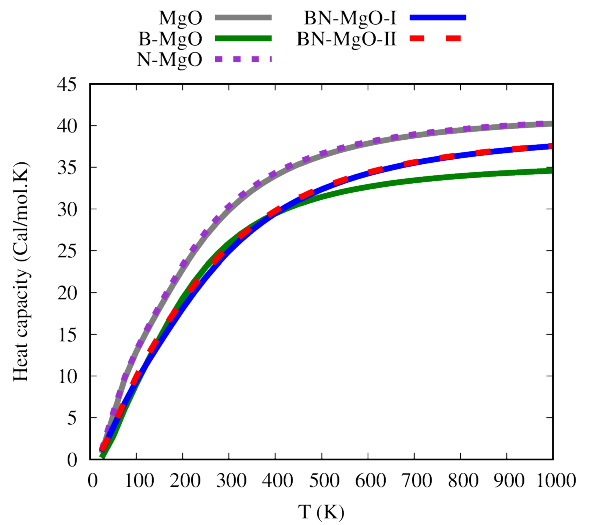}
	\caption{Heat capacity of the pure and BN-codopant MgO monolayers.}
	\label{fig06}
\end{figure}

The heat capacity of a pure and BN-codoped MgO monolayers is shown in \fig{fig06}.
As is expected, the heat capacity increases with temperature and becomes almost constant at
high values of temperature, $T>600$~K.
It is clear that the B and the N dopant atoms reduce the heat capacity over the entire range of
temperature. The predicted trend for the heat capacity of the monolayers is consistent with
the classical theory, that expects higher heat capacity for the systems with stronger
bonds \cite{MORTAZAVI2021100257}. The chemical bonds in general become stronger as the electronegativity difference across the bond increases. The average electronegativity difference across the bonds is $2.3$ (MgO), $2.18$ (N-MgO), $1.97$ (BN-MgO-II), $1.93$ (B-MgO), and $1.91$ (BN-MgO-I).
The average electronegativity of pure MgO and N-MgO are close, and almost equal for B-MgO, BN-MgO-I, BN-MgO-II. This indicates that the average bond of MgO and N-MgO is stronger than those of B-MgO, BN-MgO-I, BN-MgO-II. We can thus see that the heat capacity of MgO and N-MgO is higher than those for B-MgO, BN-MgO-I, BN-MgO-II indicating that the BN-codopant reduces the heat capacity.

The heat capacity can also be explained by the bond dissociation energy, which is the energy required for an endothermic process \cite{oxtoby2015principles}. The bond dissociation energy increases as the difference in the electronegativities of the bonded atoms increases. The higher difference in the electronegativities gives the higher endothermic process resulting in higher heat capacity.

\subsection{Optical properties}

The optical response of pure and BN-codoped MgO monolayers to electromagnetic radiation
are studied here using the random phase approximation (RPA).
In order to achieve this, we have computed various optical parameters such as the imaginary part, Im$(\varepsilon)$, and the real part, Re$(\varepsilon)$, of the complex dielectric functions,
which basically portray the absorptive and reflective character of a monolayer, respectively.
The values of Im$(\varepsilon)$ and Re$(\varepsilon)$ are then be utilized to calculate the absorption coefficient, $\alpha$, and the real part of optical conductivity, Re($\sigma_{\rm optical}$).

\begin{figure}[htb]
	\centering
	\includegraphics[width=0.5\textwidth]{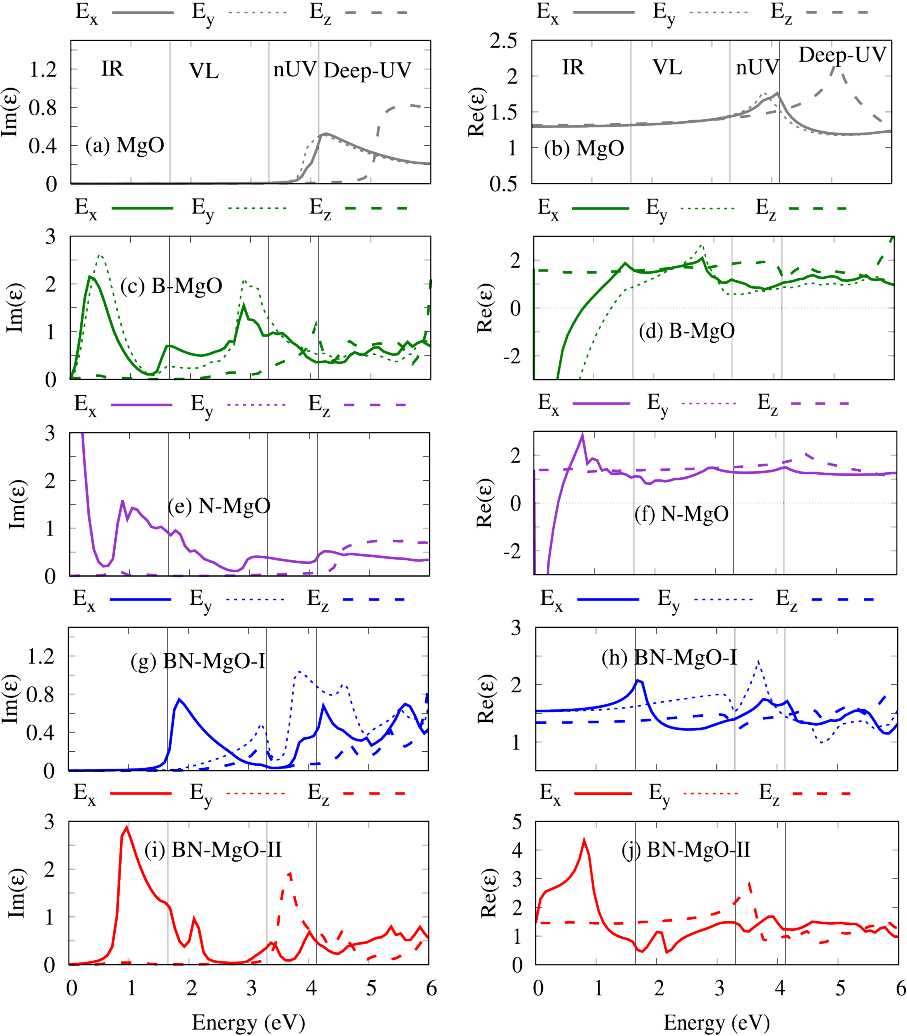}
	\caption{The Im$(\varepsilon)$ (left panel) and the Re$(\varepsilon)$ (right panel) are presented for pure and BN-codoped MgO monolayers. The solid-, dotted-, and dashed curves correspond to
		the incident light polarized along the $x$, $y$, and $z$ directions, respectively.}
	\label{fig07}
\end{figure}

In \fig{fig07}, the Im$(\varepsilon)$ (left panel) and the Re$(\varepsilon)$ (right panel) are presented for pure and BN-codoped MgO monolayers. The geometrical symmetry of pure MgO leads it to behave almost isotropicly along the $x$ and the $y$ directions for all optical parameters.
This isotropic behavior is also seen for N-MgO and BN-MgO-II monolayers, while an anisotropic property for B-MgO and BN-MgO-I is found. It indicates that the geometrical symmetry for N-MgO and BN-MgO-II is high. The high geometrical symmetry of the N-MgO monolayer is caused by the atomic weight and radii of the O and the N atoms being ``close'' to each others.
In addition, the weak attractive interaction between the B and the N atoms in BN-MgO-II causes the monolayer to be geometrically symmetric. Consequently, the isotropic character of the optical reposes of these two monolayers is established.

In contrast, the anisotropic property in the optical response for B-MgO is related to the fact that the atomic weight and radius of the O atoms are larger than those of the B atoms, as well as, the strong attractive interaction between the B and the N atoms in BN-MgO-II leads to the breaking of the geometrically symmetry and thus an anisotropic property of the monolayer is observed.

The the Im$(\varepsilon)$ spectra of MgO, BN-MgO-I, and BN-MgO-II for all directions for the polarized light confirms the semiconducting properties of these monolayers as the first peak starts from  $3.45$, $1.6$, and $0.55$~eV, respectively, indicating their band gaps. Furthermore, the Im$(\varepsilon)$ of N-MgO and B-MgO shows metallic character in the $x$- and the $y$-polarization of the incident light.

The Re$(\varepsilon)$ spectra indicate that the values of static dielectric constants, Re($\varepsilon(0)$), are $1.29$, $1.54$, $1.45$ for MgO, BN-MgO-I, and BN-MgO-II, respectively, in the $x$- and the $y$-polarization of the incident light, while $1.32$, $1.34$, and $1.45$ are found for the $z$-polarization. We find that the value of Re($\varepsilon(0)$) is related to the band gaps of the monolayers. It has been reported that that Re($\varepsilon(0)$) is inversely proportional to the band gap, Re($\varepsilon(0)$)$= 1/E_{\rm g}$ \cite{PhysRev.128.2093}. This is the reason for lower value of Re($\varepsilon(0)$) for MgO comparing to BN-MgO-I, and BN-MgO-II.

\begin{figure}[htb]
	\centering
	\includegraphics[width=0.5\textwidth]{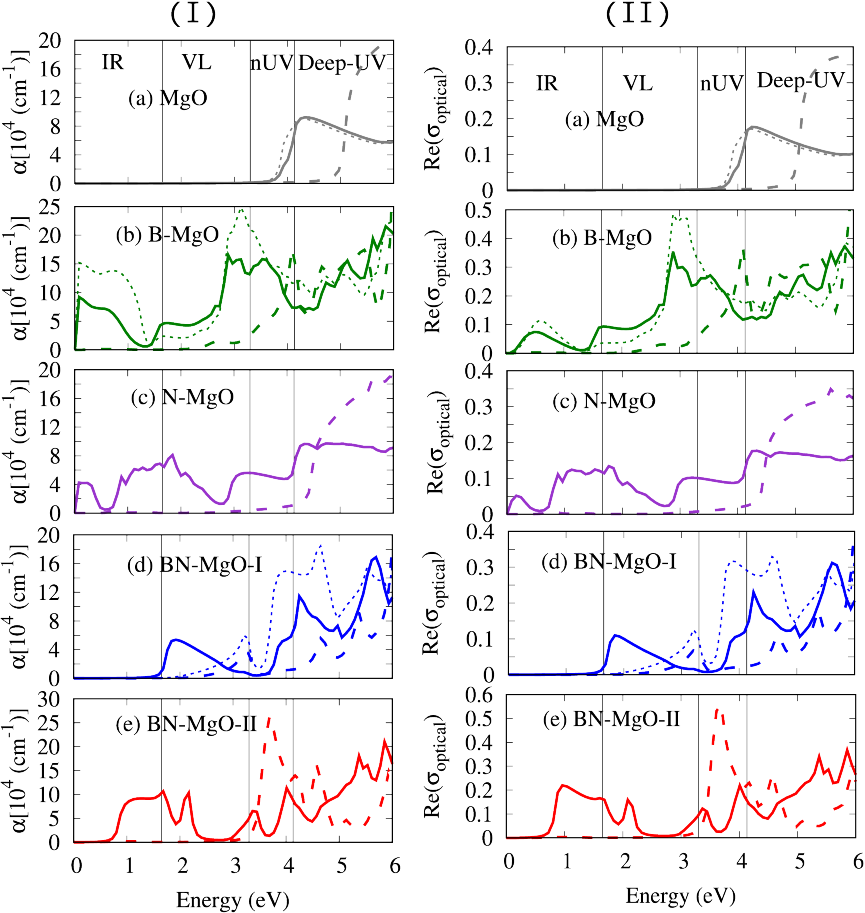}
	\caption{The absorption coefficient (I), and real part of optical conductivity (II) of the pure and BN-codoped MgO monolayers.}
	\label{fig08}
\end{figure}

Next, we study the absorption spectra, $\alpha$, and the optical conductivity, Re($\sigma_{\rm optical}$), of the monolayers shown in (I), and (II) of \fig{fig08}, respectively.
The $\alpha$ and Re($\sigma_{\rm optical}$) spectra confirm the optical band gap of the semiconducting character of the monolayers. Due to the wide band gap of MgO, the first peak in these spectra is observed in the Deep-UV region for all polarizations of the incident light. This may be beneficial for the applications in deep-UV communications, and for UV photodetectors \cite{he20191}.

The first intense peak is shifted to a lower energy of visible light region in BN-MgO-I, and the infrared region in BN-MgO-II. This red shifted peak is due to the BN-dopant atoms and important for devices working in the VL region. In addition, the isotropic and anisotropic properties are also seen here originating from the influences of the atomic weight and radius, and the attractive force between the dopant atoms, as is explained above.

\section{Conclusion}\label{conclusion}

In the present work, DFT calculations have been performed to investigate the electronic, the thermal, and
the optical properties of MgO monolayers with BN-codopant atoms.
Differently BN-copdoped MgO nanosheets have been optimized showing different strength of the inter-atomic attractive interactions between the B and the N atoms. All the pure and BN-codoped MgO monolayers with and without dopant atoms are found to exhibit planar structures.
It was seen from the partial density of state that Mg ($3s$) and O ($2p$) states contribute to the conduction band minima and the valence band maxima of pure MgO, respectively, resulting in a wide band gap.
The impurity states of the N ($2p$) and the B ($2p$) atoms contribute strongly to the valence band minima resulting in the reduction of the band gap of MgO nanosheets with BN-codopant atoms.
The reduced band gaps of the BN-codoped MgO monolayers enlarge the static values of the dielectric constant and shift the main intensity peak of the optical conductivity to the VL from the Deep-UV region.
Furthermore, the band gap reduction decreases the thermal characteristics of the monolayers. Therefore, our study of MgO nanosheets with B and N dopant atoms should be interesting for possible spin- and optoelectronic applications utilizing these materials.

\section{Acknowledgment}
This work was financially supported by the University of Sulaimani and
the Research center of Komar University of Science and Technology.
The computations were performed on resources provided by the Division of Computational
Nanoscience at the University of Sulaimani.

\section{Conflict of Interest}
The authors declare that they have no conflicts of financial interest.

\section{Compliance with Ethical Standards}
This article does not contain any studies involving human participants performed by any of the authors.

\section{Research data policy and Data Availability Statements}
The datasets generated during and/or analysed during the current study are available from the corresponding author on reasonable request.

\section{Author Contribution}

The authors confirm contribution to the paper as follows:
study conception and design: Nzar Rauf Abdullah, Vidar Gudmundsson;
data collection: Nzar Rauf Abdullah, and Yousif Hussein Azeez;
analysis and interpretation of results: Nzar Rauf Abdullah, Botan Jawdat Abdullah, Yousif Hussein Azeez, Chi-Shung Tang, and Vidar Gudmundsson;
draft manuscript preparation: Nzar Rauf Abdullah, Botan Jawdat Abdullah.
All authors reviewed the results and approved the final version of the manuscript.



\end{document}